\documentclass[journal]{IEEEtran}
\ifCLASSINFOpdf
   \usepackage[pdftex]{graphicx}
\else
\fi
\usepackage{amssymb}
\usepackage{multirow}
\usepackage{subfigure}
\usepackage{float}
\usepackage{epstopdf}
\usepackage{amsmath}
\usepackage{mathrsfs}
\usepackage{algorithmic}
\usepackage{algorithm}

\usepackage{array}
\begin{document}
%
\title{CM-GANs: Cross-modal Generative Adversarial Networks for Common Representation Learning}
%
%
%

\author{Yuxin Peng, Jinwei Qi and Yuxin Yuan
	\thanks{This work was supported by National Natural Science Foundation of China under Grants 61771025 and 61532005.}
	\thanks{The authors are with the Institute of Computer Science and Technology,	Peking University, Beijing 100871, China. Corresponding author: Yuxin Peng	(e-mail: pengyuxin@pku.edu.cn).}
}
%
%

\markboth{IEEE TRANSACTIONS ON MULTIMEDIA}%
{Shell \MakeLowercase{\textit{et al.}}: Bare Demo of IEEEtran.cls for IEEE Journals}
%



\maketitle

\begin{abstract}
It is known that the inconsistent distribution and representation of different modalities, such as image and text, cause the heterogeneity gap, which makes it very challenging to correlate such heterogeneous data. 
Recently, generative adversarial networks (GANs) have been proposed and shown its strong ability of modeling data distribution and learning discriminative representation,
and most of the existing GANs-based works mainly focus on the unidirectional generative problem to generate new data such as image synthesis.
While we have completely different goal, which aims to effectively correlate existing large-scale heterogeneous data of different modalities, by utilizing the power of GANs to model the cross-modal joint distribution. 
Thus, in this paper we propose Cross-modal Generative Adversarial Networks (CM-GANs) to learn discriminative common representation for bridging the heterogeneity gap. The main contributions can be summarized as follows:
(1) \textit{Cross-modal GANs architecture} is proposed to model the joint distribution over the data of different modalities. The inter-modality and intra-modality correlation can be explored simultaneously in generative and discriminative models. Both of them beat each other to promote cross-modal correlation learning.
(2) \textit{Cross-modal convolutional autoencoders with weight-sharing constraint} are proposed to form the generative model. They can not only exploit the cross-modal correlation for learning the common representation, but also preserve the reconstruction information for capturing the semantic consistency within each modality.
(3) \textit{Cross-modal adversarial mechanism} is proposed, which utilizes two kinds of discriminative models to simultaneously conduct intra-modality and inter-modality discrimination. They can mutually boost to make the generated common representation more discriminative by adversarial training process. 
To the best of our knowledge, our proposed CM-GANs approach is the first to utilize GANs to perform cross-modal common representation learning, by which the heterogeneous data can be effectively correlated. Extensive experiments are conducted to verify the performance of our proposed approach on cross-modal retrieval paradigm, compared with 10 state-of-the-art methods on 3 cross-modal datasets.

\end{abstract}

\begin{IEEEkeywords}
Generative adversarial network, cross-modal adversarial mechanism, common representation learning, cross-modal retrieval.
\end{IEEEkeywords}

%
\IEEEpeerreviewmaketitle

\section{Introduction}

\IEEEPARstart{M}{ultimedia} data with different modalities, including image, video, text, audio and so on, is now mixed together and represents comprehensive knowledge to perceive the real world. The research of cognitive science indicates that in human brain, the cognition of outside world is through the fusion of multiple sensory organs \cite{mcgurk1976hearing}. However, there exists the heterogeneity gap, which makes it quite difficult for artificial intelligence to simulate the above human cognitive process, because multimedia data with various modalities in Internet has huge quantity, but inconsistent distribution and representation. 

\begin{figure*}[!t]
	\centering
	\includegraphics[width=0.94\textwidth]{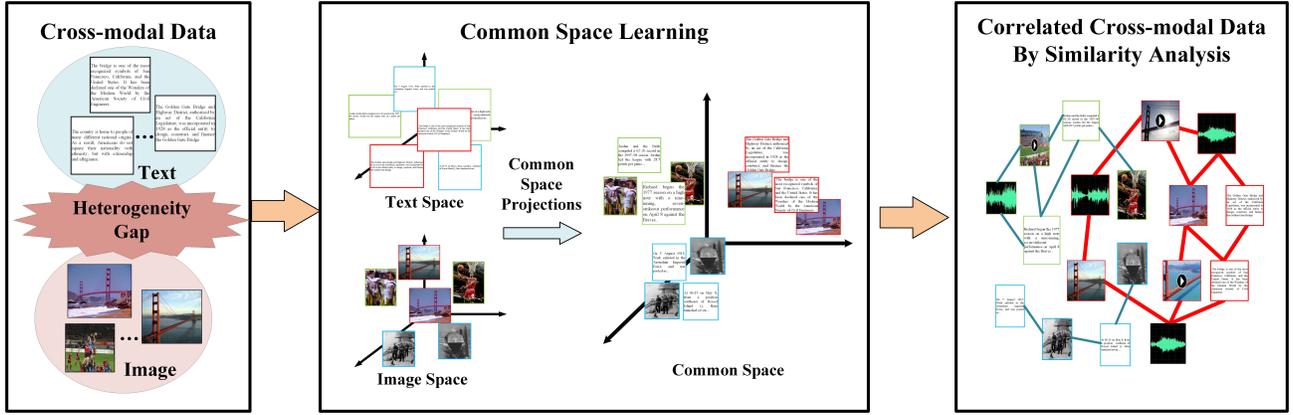}
	\caption{Illustrations of mainstream framework for constructing the cross-modal correlation learning, which aims to project the heterogeneous data of different modalities from their own feature spaces into one common space, where similarity measurement can be directly adopted to establish correlation on the cross-modal data.
	}
	\label{fig_CommonSpace}
\end{figure*}

Naturally, cross-modal correlation exists among the heterogeneous data of different modalities to describe specific kinds of statistical dependencies \cite{peng2017cross}. For example, for the image and textual descriptions coexisting in one web page, they may be intrinsically correlated from content and share a certain level of semantic consistency. 
Therefore, it is necessary to automatically exploit and understand such latent correlation across the data of different modalities, and further construct metrics on them to measure how they are semantically relevant. For addressing the above issues, an intuitive idea is to model the joint distribution over the data of different modalities to learn the common representation, which can form a commonly shared space where heterogeneous data is mapped into. Thus, the similarities between them can be directly computed by adopting common distance metrics. Figure \ref{fig_CommonSpace} shows an illustration of the above framework. In this way, the heterogeneity gap among the data of different modalities can be reduced, so that the heterogeneous data can be correlated together more easily to realize various practical application, such as cross-modal retrieval \cite{peng2017overview}, where the data of different modalities can be retrieved at the same time by a query of any modality flexibly.

Following the above idea, some methods \cite{yang2008harmonizing,DBLP:journals/tmm/ZhuangYW08,zhang2016cross} have been proposed to deal with the heterogeneity gap and learn the common representation by modeling the cross-modal correlation, so that the similarities between different modalities can be measured directly. 
These existing methods can be divided into two major categories according to their different models as follows: 
The first is in traditional framework, which attempts to learn mapping matrices for the data of different modalities by optimizing the statistical values, so as to project them into one common space. Canonical Correlation Analysis (CCA) \cite{RasiwasiaMM10SemanticCCA} is one of the representative works, which has many extensions such as \cite{DBLP:journals/ijcv/GongKIL14,DBLP:journals/neco/HardoonSS04}. The second kind of methods \cite{DBLP:conf/ijcai/PengHQ16,feng12014cross,DBLP:journals/tmm/PangZN15} utilize the strong learning ability of deep neural network (DNN) to construct multilayer network, and most of them aim to minimize the correlation learning error across different modalities for the common representation learning.

Recently, generative adversarial networks (GANs) \cite{goodfellow2014generative} have been proposed to estimate a generative model by an adversarial training process. The basic model of GANs consists of two components, namely a generative model \textit{G} and a discriminative model \textit{D}. The generative model aims to capture the data distribution, while the discriminative model attempts to discriminate whether the input sample comes from real data or is generated from \textit{G}. Furthermore, an adversarial training strategy is adopted to train these two models simultaneously, which makes them compete with each other for mutual promotion to learn better representation of the input data. Inspired by the recent progress of GANs, researchers attempt to apply GANs into computer vision areas, such as image synthesis \cite{radford2015unsupervised}, video prediction \cite{finn2016unsupervised} and object detection \cite{li2017perceptual}. 

Due to the strong ability of GANs in modeling data distribution and learning discriminative representation, it is a natural solution that GANs can be utilized for modeling the joint distribution over the heterogeneous data of different modalities, which aims to learn the common representation and boost the cross-modal correlation learning.
However, most of the existing GANs-based works only focus on the unidirectional generative problem to generate new data for some specific applications, 
such as image synthesis to generate certain image by a noise input \cite{radford2015unsupervised}, side information \cite{mirza2014conditional} or text description \cite{reed2016generative}. Their main purpose is to generate new data of single modality, which cannot effectively establish correlation on multimodal data with heterogeneous distribution. 
Different from the existing works, we aim to utilize GANs for establishing correlation on the existing large-scale heterogeneous data of different modalities by common representation generation, which is a completely different goal to model the joint distribution over the multimodal input.

For addressing the above issues, we propose Cross-modal Generative Adversarial Networks (CM-GANs), which aims to learn discriminative common representation with multi-pathway GANs to bridge the gap between different modalities. The main contributions can be summarized as follows.
\begin{itemize}
	\item{
		\textbf{\textit{Cross-modal GANs architecture}} is proposed to deal with the heterogeneity gap between different modalities, which can effectively model the joint distribution over the heterogeneous data simultaneously. The generative model learns to fit the joint distribution by modeling inter-modality correlation and intra-modality reconstruction information, while the discriminative model leans to judge the relevance both within the same modality and between different modalities. Both generative and discriminative models beat each other with a minimax game for better cross-modal correlation learning.
	} 
	\item{
		\textbf{\textit{Cross-modal convolutional autoencoders with weight-sharing constraint}} are proposed to form the two parallel generative models. Specifically, the encoder layers contain convolutional neural network to learn high-level semantic information for each modality, and also exploit the cross-modal correlation by the weight-sharing constraint for learning the common representation. While the decoder layers aim to model the reconstruction information, which can preserve semantic consistency within each modality.
	} 
	\item{
		\textbf{\textit{Cross-modal adversarial mechanism}} is proposed to perform a novel adversarial training strategy in cross-modal scenario, which utilizes two kinds discriminative models to simultaneously conduct inter-modality and intra-modality discrimination. Specifically, the inter-modality discrimination aims to discriminate the generated common representation from which modality, while the intra-modality discrimination tends to discriminate the generated reconstruction representation from the original input, which can mutually boost to force the generative models to learn more discriminative common representation by adversarial training process.
	} 
\end{itemize}

\begin{figure}[!t]
	\centering
	\includegraphics[width=0.49\textwidth]{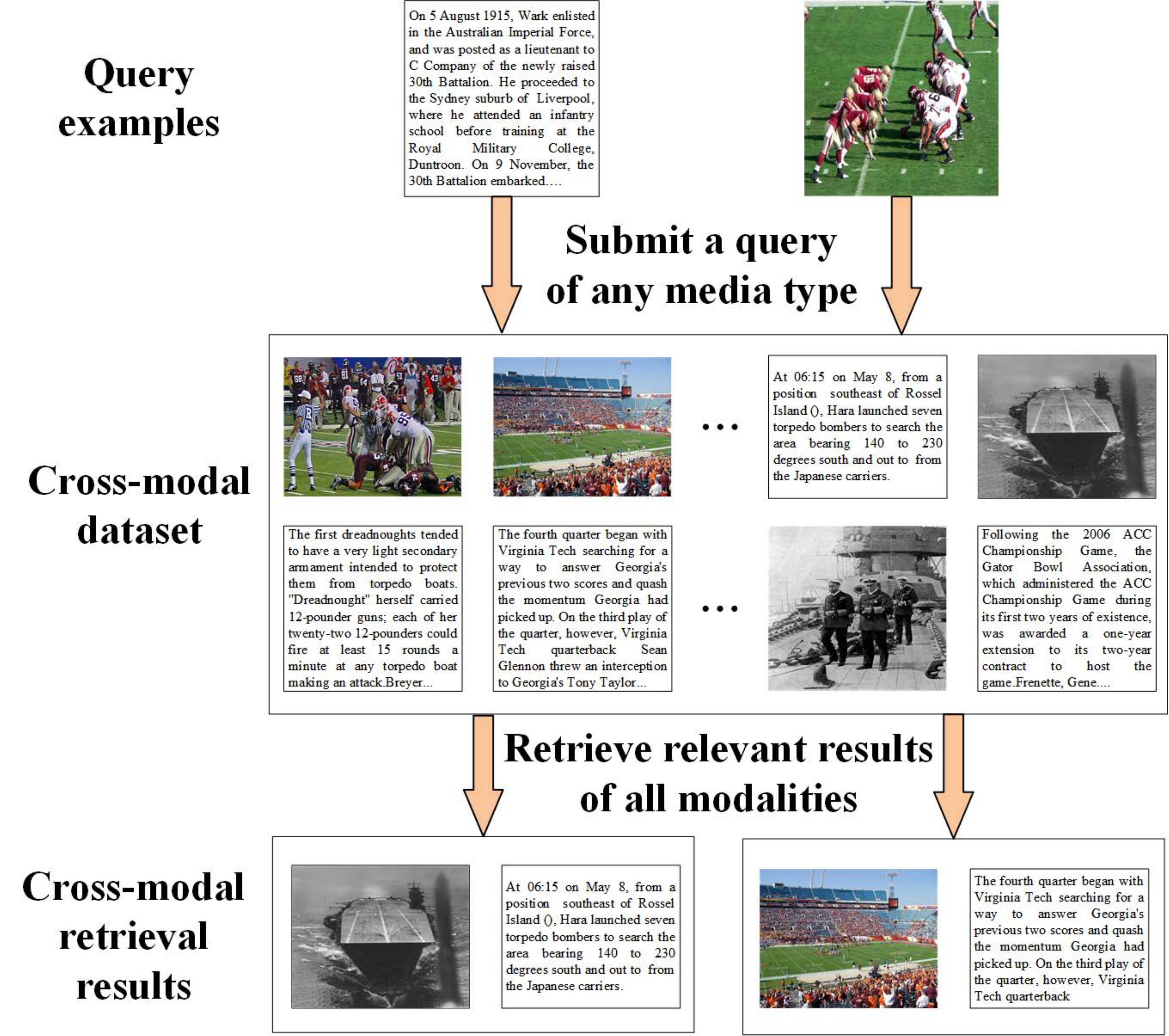}
	\caption{An example of cross-modal retrieval with image and text, which can	present retrieval results with different modalities by a query of any modality.}
	\label{fig_cross_media}
\end{figure}

To the best of our knowledge, our proposed CM-GANs approach is the first to utilize GANs to perform cross-modal common representation learning. 
With learned common representation, heterogeneous data can be correlated by common distance metric. We conduct extensive experiments on cross-modal retrieval paradigm, to evaluate the performance of cross-modal correlation, which aims to retrieve the relevant results across different modalities by distance metric on the learned common representation, as shown in Figure \ref{fig_cross_media}. Comprehensive experimental results show the effectiveness of our proposed approach, where our proposed approach achieves the best retrieval accuracy compared with 10 state-of-the-art cross-modal retrieval methods on 3 widely-used datasets: Wikipedia, Pascal Sentence and our constrcuted large-scale XmediaNet datasets.

The rest of this paper is organized as follows: We first briefly introduce the related works on cross-modal correlation learning methods as well as existing GANs-based methods in Section II. Section III presents our proposed CM-GANs approach. Section IV introduces the experiments of cross-modal retrieval conducted on 3 cross-modal datasets with the result analyses. Finally Section V concludes this paper.

\section{Related Works}

In this section, the related works are briefly reviewed from the following two aspects: cross-modal correlation learning methods, and representative works based on GANs.

\subsection{Cross-modal Correlation Learning Methods}

For bridging the heterogeneity gap between different modalities, there are some methods proposed to conduct cross-modal correlation learning, which mainly aim to learn the common representation and correlate the heterogeneous data by distance metric. We briefly introduce the representative methods of cross-modal correlation learning with the following two categories, namely \textit{traditional methods} and \textit{deep learning based methods}.

\textit{Traditional methods} mainly learn linear projection to maximize the correlation between the pairwise data of different modalities, which project the feature of different modalities into one common space to generate common representation. One class of methods attempt to optimize the statistical values to perform statistical correlation analysis. A representative method is to adopt canonical correlation analysis (CCA) \cite{HotelingBiometrika36RelationBetweenTwoVariates} to construct a lower-dimensional common space, which has many extensions, such as adopting kernel function \cite{DBLP:journals/neco/HardoonSS04}, integrating semantic category labels \cite{RasiwasiaMM10SemanticCCA}, taking high-level semantics as the third view to perform multi-view CCA \cite{DBLP:journals/ijcv/GongKIL14}, and considering the semantic information in the form of multi-label annotations \cite{DBLP:conf/iccv/RanjanRJ15}. Besides, cross-modal factor analysis (CFA) \cite{LiMM03CFA}, which is similar to CCA, minimizes the Frobenius norm between the pairwise data to learn the projections for common space. Another class of methods integrate graph regularization into the cross-modal correlation learning, namely to construct graphs for correlating the data of different modalities in the learned common space. For example, Zhai et al. \cite{ZhaiAAAI2013JGRHML} propose joint graph regularized heterogeneous metric learning (JGRHML) method, which adopts both metric learning and graph regularization, and they further integrate semi-supervised information to propose joint representation learning (JRL) \cite{ZhaiTCSVT2014JRL}. Wang et al. \cite{DBLP:journals/pami/WangHWWT16} also adopt graph regularization to simultaneously preserve the inter-modality and intra-modality correlation.

As for the \textit{deep learning based methods}, with the strong power of non-linear correlation modeling, deep neural network has made great progress in numerous single-modal problems such as object detection \cite{DBLP:conf/nips/RenHGS15} and image classification \cite{ImageNet2012}. It has also been utilized to model the cross-modal correlation. Ngiam et al. \cite{ngiam32011multimodal} propose bimodal autoencoder, which is an extension of restricted Boltzmann machine (RBM), to model the cross-modal correlation at the shared layer, and followed by some similar network structures such as \cite{kim2012learning}-\cite{DBLP:conf/ijcai/WangCO015}. Multimodal deep belief network \cite{srivastava2012learning} is proposed to model the distribution over the data of different modalities and learn the cross-modal correlation by a joint RBM. Feng et al. \cite{feng12014cross} propose correspondence autoencoder (Corr-AE), which jointly models the cross-modal correlation and reconstruction information. Deep canonical correlation analysis (DCCA) \cite{DBLP:conf/icml/AndrewABL13,DBLP:conf/cvpr/YanM15} attempts to combine deep network with CCA. The above methods mainly contain two subnetworks linked at the joint layer for correlating the data of different modalities. Furthermore, cross-modal multiple deep networks (CMDN) are proposed \cite{DBLP:conf/ijcai/PengHQ16} to construct hierarchical network structure for both inter-modality and intra-modality modeling. Cross-modal correlation learning (CCL) \cite{peng2017ccl} method is further proposed to integrate fine-grained information as well as multi-task learning strategy for better performance.

\begin{figure*}[t]
	\centering
	\begin{minipage}[c]{0.97\linewidth}
		\centering
		\includegraphics[width=1.0\textwidth]{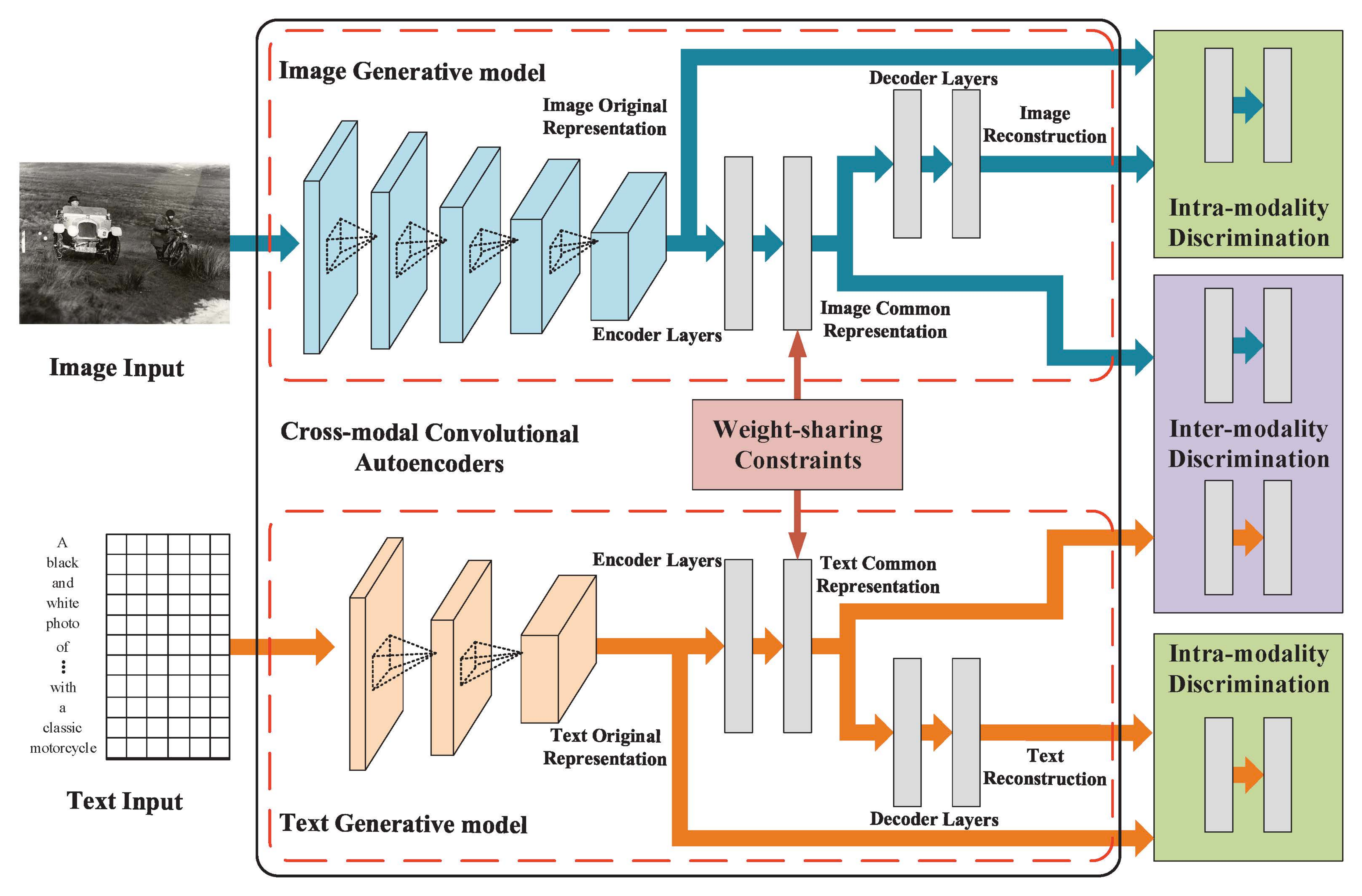}
	\end{minipage}%
	\caption{An overview of our CM-GANs approach with two main components. First, cross-modal convolutional autoencoders form the generative model to exploit both cross-modal correlation as well as reconstruction information. Second, two kinds of discriminative models conduct intra-modality discrimination and inter-modality discrimination simultaneously to perform cross-modal adversarial training.}\label{fig:network}
\end{figure*}

\subsection{Generative Adversarial Networks}

Since generative adversarial networks (GANs) have been proposed by Goodfellow et al. \cite{goodfellow2014generative} in 2014, a series of GANs-based methods have arisen for a wide variety of problems. The original GANs consist of a generative model \textit{G} and a discriminative model \textit{D}, which aim to learn a generative model for capturing the distribution over real data with an adversarial discriminative model, in order to discriminate between real data and generated fake data. Specifically, $D$ and $G$ play the minimax game on $V(G,D)$ as follows.
\begin{align}
\underset{G}{\textrm{min}}\ \underset{D}{\textrm{max}}\ V(G,D)=&\mathbb{E}_{x\sim p_{data}(x)}[\textrm{log}D(x)]+ \notag \\
&\mathbb{E}_{x\sim p_{z}(z)}[\textrm{log}(1-D(G(z)))]
\end{align}
where $x$ denotes the real data and $z$ is the noise input, and this minimax game has a global optimum when $p_g=p_{data}$. Furthermore, Mirza et al. \cite{mirza2014conditional} propose conditional GANs (cGAN) to condition the generate data with side information instead of uncontrollable noise input.

Most of the existing GANs-based works focus on the generative problem to generate new data, and they are mainly developed for some specific applications. One of the most popular applications is image synthesis to generate natural images. Radford et al. \cite{radford2015unsupervised} propose deep convolutional generative adversarial networks (DCGANs) to generate images from a noise input by using deconvolutions. Denton et al. \cite{denton2015deep} utilize conditional GANs and propose a Laplacian pyramid framework with cascade of convolutional networks to generate images in a coarse-to-fine fashion. Besides image generation, Ledig et al. \cite{ledig2016photo} propose super-resolution generative adversarial network (SRGAN), which designs a perceptual loss function consisting of an adversarial loss and a content loss. Wang et al. \cite{wang2016generative} propose style and structure generative adversarial network (S$^2$-GAN) to conduct style transfer from normal maps to the realistic images. Li et al. \cite{li2017perceptual} propose perceptual GAN that performs small object detection by transferring poor representation of small object to super-resolved one to improve the detection performance.

The methods mentioned above cannot handle the multimedia data with heterogeneous distribution. Recently, there are some works proposed to explore the multimedia application. Odena et al. \cite{odena2016conditional} attempt to generate images conditioned on class labels, which forms auxiliary classifier GAN (AC-GAN). Reed et al. \cite{reed2016generative} utilize GANs to translate visual concepts from sentences to images. They further propose the generative adversarial what-where network (GAWWN) \cite{reed2016learning} to generate images by giving the description on what content to draw in which location. Furthermore, Zhang et al. \cite{zhang2016stackgan} adopt StackGAN to synthesize photo-realistic images from text descriptions, which can generate higher resolution images than the prior work \cite{reed2016generative}. 

However, the aforementioned works still have limited flexibility, because they only address the generative problem from one modality to another through one pathway network structure unidirectionally. They cannot model the joint distribution over the multimodal input to correlate the large-scale heterogeneous data. Inspired by GANs' strong ability in modeling data distribution and learning discriminative representation, we utilize GANs for modeling the joint distribution over the data of different modalities to learn the common representation, which aims to further construct correlation on the large-scale heterogeneous data across various modalities.

\section{Our CM-GANs Approach}

The overall framework of our proposed CM-GANs approach is shown in Figure \ref{fig:network}. For the generative model, \textit{cross-modal convolutional autoencoders} are adopted to generate the common representation by exploiting the cross-modal correlation with \textit{weight-sharing constraint}, and also generate the reconstruction representation aiming to preserve the semantic consistency within each modality. For the discriminative model, two kinds of discriminative models are designed with \textit{inter-modality and intra-modality discrimination}, which can make discrimination on both the generated common representation as well as the generated reconstruction representation for mutually boosting. The above two models are trained together with \textit{cross-modal adversarial mechanism} for learning more discriminative common representation to correlate heterogeneous data of different modalities.

\subsection{Notation}

The formal definition is first introduced. We aim to conduct correlation learning on the multimodal dataset, which consists of two modalities, namely $I$ as image and $T$ as text. The multimodal dataset is represented as $D=\left \{D_{tr},D_{te}\right \}$, where $D_{tr}$ denotes the training data and testing data is $D_{te}$. Specifically, $D_{tr}=\{I_{tr},T_{tr}\}$, where $I_{tr}=\{i_p\}_{p=1}^{n_{tr}}$ and $T_{tr}=\{t_p\}_{p=1}^{n_{tr}}$. $i_p$ and $t_p$ are the $p$-th instance of image and text, and totally $n_{tr}$ instances of each modality are in the training set. Furthermore, there are semantic category labels $\{c_p^I\}_{p=1}^{n_{tr}}$ and $\{c_p^T\}_{p=1}^{n_{tr}}$ for each image and text instance respectively. As for the testing set denoted as $D_{te}=\{I_{te},T_{te}\}$, there are $n_{te}$ instances for each modality including $I_{te}=\{i_q\}_{q=1}^{n_{te}}$ and $T_{te}=\{t_q\}_{q=1}^{n_{te}}$. 

Our goal is to learn the common representation for each image or text instance so as to calculate cross-modal similarities between different modalities, which can correlate the heterogeneous data. For further evaluating the effectiveness of the learned common representation, cross-modal retrieval is conducted based on the common representation, which aims to retrieve the relevant text $t_q$ from $T_{te}$ by giving an image query $i_q$ from $I_{te}$, and vice versa to retrieve image by a query of text. In the following subsections, first our proposed network architecture is introduced, then followed by the objective functions of the proposed CM-GANs, and finally the training procedure of our model is presented.

\subsection{Cross-modal GANs architecture}

As shown in Figure \ref{fig:network}, we introduce the detailed network architectures of the generative model and discriminative model in our proposed CM-GANs approach respectively as follows.

\subsubsection{\textbf{Generative model}} We design cross-modal convolutional autoencoders to form two parallel generative models for each modality respectively, denoted as $G_I$ for image and $G_T$ for text, which can be divided into encoder layers $G_{Ienc}$ and $G_{Tenc}$ as well as decoder layers $G_{Idec}$ and $G_{Tdec}$.
The encoder layers contain convolutional neural network to learn high-level semantic information for each modality, followed by several fully-connected layers which is linked at the last one with weight-sharing and semantic constraints to exploit cross-modal correlation for the common representation learning. While the decoder layers aim to reconstruct the high-level semantic representation obtained from the convolutional neural network ahead, which can preserve semantic consistency within each modality.

For image data, each input image $i_p$ is first resized into $256\times 256$, and then fed into the convolutional neural network to exploit the high-level semantic information. The encoder layers have the following two subnetworks: The convolutional layers have the same configuration as the 19-layer VGG-Net \cite{DBLP:journals/corr/SimonyanZ14a}, which is pre-trained on the ImageNet and fine-tuned on the training image data $I_{tr}$. We generate 4,096 dimensional feature vector from fc7 layer as the original high-level semantic representation for image, denoted as $h_p^i$. Then, several additional fully-connected layers conduct the common representation learning, where the learned common representation for image is denoted as $s_p^i$. The decoder layers have a relatively simple structure, which contain several fully-connected layers to generate the reconstruction representation $r_p^i$ from $s_p^i$, in order to reconstruct $h_p^i$ to preserve semantic consistency of image.

For text data, assuming that input text instance $t_p$ consists of $n$ words, each word is represented as a $k$-dimensional feature vector, which is extracted by Word2Vec model \cite{DBLP:conf/nips/MikolovSCCD13} pre-trained on billions of words in Google News. Thus, the input text instance can be represented as an $n\times k$ matrix. The encoder layers also have the following two subnetworks: The convolutional layers on the input matrix have the same configuration with \cite{DBLP:conf/emnlp/Kim14} to generate the original high-level semantic representation for text, denoted as $h_p^t$. Similarly, there follow several additional fully-connected layers to learn text common representation, denoted as $s_p^t$. The decoder layers aim to preserve semantic consistency of text by reconstructing $h_p^t$ with the generated reconstruction representation $r_p^t$, which is also made up of fully-connected layers.

For the weight-sharing and semantic constraints, we aim to correlate the generated common representation of each modality. Specifically, the weights of last few layers of the image and text encoders are shared, which are responsible for generating the common representation of image and text, with the intuition that common representation for a pair of corresponding image and text should be as similar as possible. Furthermore, the weight-sharing layers are followed by a softmax layer for further exploiting the semantic consistency between different modalities. Thus, the cross-modal correlation can be fully modeled to generate more discriminative common representation.

\subsubsection{\textbf{Discriminative model}} We adopt two kinds discriminative models to simultaneously conduct intra-modality and inter-modality discrimination. Specifically, the intra-modality discrimination aims to discriminate the generated reconstruction representation from the original input, while the inter-modality discrimination tends to discriminate the generated common representation from image or text modality.

For the intra-modality discriminative model, it consists of two subnetworks for image and text, denoted as $D_I$ and $D_T$ respectively. Each of them is made up of several fully-connected layers, which takes original high-level semantic representation as the real data and reconstruction representation as the generated data to make discrimination.
For the inter-modality discriminative model denoted as $D_C$, a two-pathway network is also adopted, where $D_{Ci}$ is for image pathway and $D_{Ct}$ is for text pathway. Both of them aim to discriminate which modality the common representation is from, such as on image pathway to discriminate the image common representation as the real data, while its corresponding text common representation and mismatched image common representation with different category as the fake data. So as for the text pathway to discriminate the text common representation with others.



\subsection{Objective Functions}

The objective of the proposed generative model and discriminative model in CM-GANs is defined as follows.

\begin{itemize}
	\item \textit{\textbf{Generative model}}: As shown in the middle of Figure \ref{fig:network}, $G_I$ and $G_T$ for image and text respectively, each of which generates three kinds of representations, namely original representation and common representation from $G_{Ienc}$ or $G_{Tenc}$, also the reconstruction representation from $G_{Idec}$ or $G_{Tdec}$, such as $h_p^i$, $s_p^i$ and $r_p^i$ for image instance $i_p$. The goal of generative model is to fit the joint distribution by modeling both inter-modality correlation and intra-modality reconstruction information.
	\item \textit{\textbf{Discriminative model}}: As shown in the right of Figure \ref{fig:network}, $D_I$ and $D_T$ for intra-modality discrimination, while $D_{Ci}$ and $D_{Ct}$ for inter-modality discrimination. For the intra-modality discrimination, $D_I$ aims to distinguish the real image data $h_p^i$ with the generated reconstruction data $r_p^i$, and $D_T$ is similar for text. For the inter-modality discrimination, $D_{Ci}$ tries to discriminate the image common representation $s_p^i$ as the real data with both text common representation $s_p^t$ and the common representation of mismatched image $\hat{s_p^i}$ as fake data. Each of them is concatenated with their corresponding original image representation $h_p^i$ and $\hat{h_p^i}$ of the mismatched one for better discrimination. $D_{Ct}$ is similar to discriminate the text common representation $s_p^t$ with the fake data of $s_p^i$ and $\hat{s_p^t}$.
\end{itemize}

With the above definitions, the generative model and discriminative model can beat each other with a minimax game, and our CM-GANs can be trained by jointly solving the learning problem of two parallel GANs.
\begin{align}
\underset{G_I,G_T}{\textrm{min}}\ \underset{D_I,D_T,D_{Ci},D_{Ct}}{\textrm{max}}\ &\mathcal{L}_{GAN_1}(G_I,G_T,D_I,D_T) \notag \\
+&\mathcal{L}_{GAN_2}(G_I,G_T,D_{Ci},D_{Ct})
\label{equ_gan0}
\end{align}
The generative model aims to learn more similar common representation for the instances between different modalities with the same category, as well as more close reconstruction representation within each modality to fool the discriminative model, while the discriminative model tries the distinguish each of them to conduct intra-modality discrimination with $\mathcal{L}_{GAN_1}$ and inter-modality discrimination with $\mathcal{L}_{GAN_2}$. 
The objective functions are given as follows.
\begin{align}
\mathcal{L}_{GAN_1}=&\mathbb{E}_{i\sim P_i}[D_I(i)-D_I(G_{Idec}(i))] \notag \\ +&\mathbb{E}_{t\sim P_t}[D_T(t)-D_T(G_{Tdec}(t))] \label{equ_ganob1}\\
\mathcal{L}_{GAN_2}=&\mathbb{E}_{i,t\sim P_{i,t}}[D_{Ci}(G_{Ienc}(i))-\frac{1}{2}D_{Ci}(G_{Tenc}(t)) \notag \\ -&\frac{1}{2}D_{Ci}(G_{Ienc}(\hat{i}))
+D_{Ct}(G_{Tenc}(t)) \notag \\-&\frac{1}{2}D_{Ct}(G_{Ienc}(i))-\frac{1}{2}D_{Ct}(G_{Tenc}(\hat{t}))] 
\label{equ_ganob2}
\end{align}
With the above objective functions, the generative model and discriminative model can be trained iteratively to learn more discriminative common representation for different modalities.

\subsection{Cross-modal Adversarial Training Procedure}

With the defined objective functions in equations (\ref{equ_ganob1}) and (\ref{equ_ganob2}), the generative model and discriminative model are trained iteratively in an adversarial way. The parameters of generative model are fixed during the discriminative model training stage and vice versa. It should be noted that we keep the parameters of convolutional layers fixed during the training phase, for the fact that our main focus is the cross-modal correlation learning. In the following paragraphs, the optimization algorithms of these two models are presented respectively.

\subsubsection{\textbf{Optimizing discriminative model}}

For the intra-modality discrimination, as shown in Figure \ref{fig:network}, taking image pathway as an example, we first generate the original high-level representation $h_p^i$ and the reconstruction representation $r_p^i$ from the generative model. Then, the intra-modality discrimination for image aims to maximize the log-likelihood for correctly distinguishing $h_p^i$ as the real data and $r_p^i$ as the generated reconstruction data, by ascending its stochastic gradient as follows:
\begin{align}
\bigtriangledown _{\theta_{D_I}}\frac{1}{N} \sum_{p=1}^{N}[\textrm{log}(1-D_I(r_p^i))+\textrm{log}(D_I(h_p^i))]
\label{equ_gradient_di}
\end{align}
where $N$ is the number of instance in one batch. Similarly, the intra-modality discriminative model for text $D_T$ can be updated with the following equation:
\begin{align}
\bigtriangledown _{\theta_{D_T}}\frac{1}{N} \sum_{q=1}^{N}[\textrm{log}(1-D_T(r_q^t))+\textrm{log}(D_T(h_q^t))]
\label{equ_gradient_dt}
\end{align}

Next, for the inter-modality discrimination, there is also a two-pathway network for each modality. As for the image pathway, inter-modality discrimination is conducted to maximize the log-likelihood fo correctly discriminate the common representation of different modalities, specifically $s_p^i$ as the real data while the text common representation $s_p^t$ and mismatching image instance $\hat{s_p^i}$ with different categories as the fake data, which are also concatenated with their corresponding original representation $h_p^i$ or $\hat{h_p^i}$ of mismatched one for better discrimination. The stochastic gradient is calculated as follows:
\begin{align}
\label{equ_gradient_dc1}
\bigtriangledown _{\theta_{D_{Ci}}}&\frac{1}{N} \sum_{p=1}^{N}[\textrm{log}D_{Ci}(s_p^i,h_p^i)+ \\
&(\textrm{log}(1-D_{Ci}(s_p^t,h_p^i))+\textrm{log}(1-D_{Ci}(\hat{s_p^i},\hat{h_p^i})))/2] \notag
\end{align}
where $(s,h)$ means to concatenate the two representations, and so as to the following equations. Similarly, for the text pathway, the stochastic gradient can be calculated with following equation:
\begin{align}
\label{equ_gradient_dc2}
\bigtriangledown _{\theta_{D_{Ct}}}&\frac{1}{N} \sum_{q=1}^{N}[\textrm{log}D_{Ct}(s_q^t,h_q^t)+ \\
&(\textrm{log}(1-D_{Ct}(s_q^i,h_q^t))+\textrm{log}(1-D_{Ct}(\hat{s_q^t},\hat{h_q^t})))/2] \notag
\end{align}

\subsubsection{\textbf{Optimizing generative model}}

There are two generative models for image and text. The image generative model aims to minimize the object function to fit true relevance distribution, which is trained by descending its stochastic gradient with the following equation, while the discriminative model is fixed at this time.
\begin{align}
\bigtriangledown _{\theta_{G_I}}\frac{1}{N} \sum_{p=1}^{N}[\textrm{log}D_{Ct}(s_p^i,h_p^t)+\textrm{log}(D_I(r_p^i))]
\label{equ_gradient_gi}
\end{align}
where $(s_p^i,h_p^t)$ also means the concatenation of the two representation. For the text generative model, it is updated similarly by descending the stochastic gradient as follows:
\begin{align}
\bigtriangledown _{\theta_{G_T}}\frac{1}{N} \sum_{q=1}^{N}[\textrm{log}D_{Ci}(s_q^t,h_q^i)+\textrm{log}(D_T(r_q^t))]
\label{equ_gradient_gt}
\end{align}

In summary, the overall training procedure is presented in Algorithm \ref{alg_train}. It should be noted that the training procedure between the generative and discriminative models needs to be carefully balanced, for the fact that there are multiple different kinds of discriminative models to provide gradient information for inter-modality and intra-modality discrimination. Thus the generative model is trained for $K$ steps in each iteration in training stage to learn more discriminative representation. 

\begin{algorithm}
	\caption{CM-GANs training procedure}
	\label{alg_train}
	\begin{algorithmic}[1] 
		\REQUIRE  
		Image training data $I_{tr}$, text training data $T_{tr}$, batchsize $N$, the number of training steps to apply generative model $K$, learning rate $\alpha$.
		\REPEAT
		\STATE Sample matching image and text pair $\{i_p\}_{p=1}^N\subseteq I_{tr}$ and $\{t_p\}_{p=1}^N\subseteq T_{tr}$, meanwhile each of them has a mismatching instance $\hat{i_p}$ and $\hat{t_p}$.
		\STATE 
		Image pathway generation $G_I(i_p)\rightarrow (h_p^i,s_p^i,r_p^i)$.
		\STATE 
		Text pathway generation $G_T(t_p)\rightarrow (h_p^t,s_p^t,r_p^t)$.
		\STATE Update intra-modality discriminative model for image $D_I$ with ascending its stochastic gradient by equation (\ref{equ_gradient_di}).
		\STATE Update intra-modality discriminative model for text $D_T$ with ascending its stochastic gradient by equation (\ref{equ_gradient_dt}).
		\STATE Update inter-modality discriminative model $D_C$ with ascending its stochastic gradient by equations (\ref{equ_gradient_dc1}) for $D_{Ci}$ and (\ref{equ_gradient_dc2}) for $D_{Ct}$.
		\FOR {K steps}
		\STATE Sample data similarly with Step 2.
		\STATE Update the image generative model by descending its stochastic gradient by equation (\ref{equ_gradient_gi}).
		\STATE Update the text generative model by descending its stochastic gradient by equation (\ref{equ_gradient_gt}).
		\ENDFOR
		\UNTIL {CM-GANs converges}
		\RETURN Optimized CM-GANs model.
	\end{algorithmic}
\end{algorithm}

\subsection{Implementation Details}

Our proposed CM-GANs approach is implemented by Torch\footnote{http://torch.ch/}, which is widely used as a scientific computing framework. The implementation details of generative model and discriminative model are introduced respectively in the following paragraphs.

\subsubsection{Generative model}

The generative model is in the form of cross-modal convolutional autoencoders with two pathways for image and text. The convolutional layers in encoder have the same configuration with 19-layer VGG-Net \cite{DBLP:journals/corr/SimonyanZ14a} for image pathway and word CNN \cite{DBLP:conf/emnlp/Kim14} for text pathway as mentioned above. Then two fully-connected layers are adopted in each pathway, and each layer is followed by a batch normalization layer and a ReLU activation function layer. The numbers of hidden units for the two layers are both 1,024. Through the encoder layers, we can get the common representation for image and text. 
Weights of the second fully-connected layer between text and image pathway are shared to learn the correlation of different modalities. The structure of decoder is made up of two fully-connected layers on each pathway, except there is no subsequent layer after the second fully-connected layer. The dimension of the first layer is 1,024 and that of the second layer is the same with the original representation obtained by CNN. What's more, the common representations are fed into a softmax layer for the semantic constraint. 

\subsubsection{Intra-modality Discriminative model}

The discriminative model for intra-modality is a network with one fully-connected layer for each modality, which can project the input feature vector into the single-value predict score, followed by a sigmoid layer. For distinguishing the original representation of the instance and the reconstructed representation, we label the original ones with 1 and reconstructed ones with 0 during discriminative model training phase.

\subsubsection{Inter-modality Discriminative model}

The discriminative model for inter-modality is a two-pathway network, and both of them take the concatenation of the common representation and the original representation as input. Each pathway consists of two fully-connected layers. The first layer has 512 hidden units, followed by a batch normalization layer and a ReLU activation function layer. The second layer generates the single-value predicted score from the output of the first layer and feed into a sigmoid layer, which is similar with the intra-modality discriminative model. For the image pathway, the image common representation is labeled with 1, while its corresponding text representation and mismatched image common representation are labeled with 0, and vice versa for the text pathway.

\section{Experiments}

In this section, we will introduce the configurations of the experiments and show the results and analyses. Specifically, we conduct the experiments on three datasets, including two widely-used datasets, Wikipedia and Pascal Sentence datasets, and one large-scale XMediaNet dataset constructed by ourselves. We compare our approach with 10 state-of-the-art methods and 5 baseline approaches to verify the effectiveness of our approach and the contribution of each component comprehensively.

\subsection{Datasets}

The datasets used in the experiments are briefly introduced first, which are XMediaNet, Pascal Sentence and Wikipedia datasets. The first is a large-scale cross-modal dataset which is constructed by ourselves. The others are widely used in cross-modal task.

\begin{figure}[!t]
	\centering
	\includegraphics[width=0.45\textwidth]{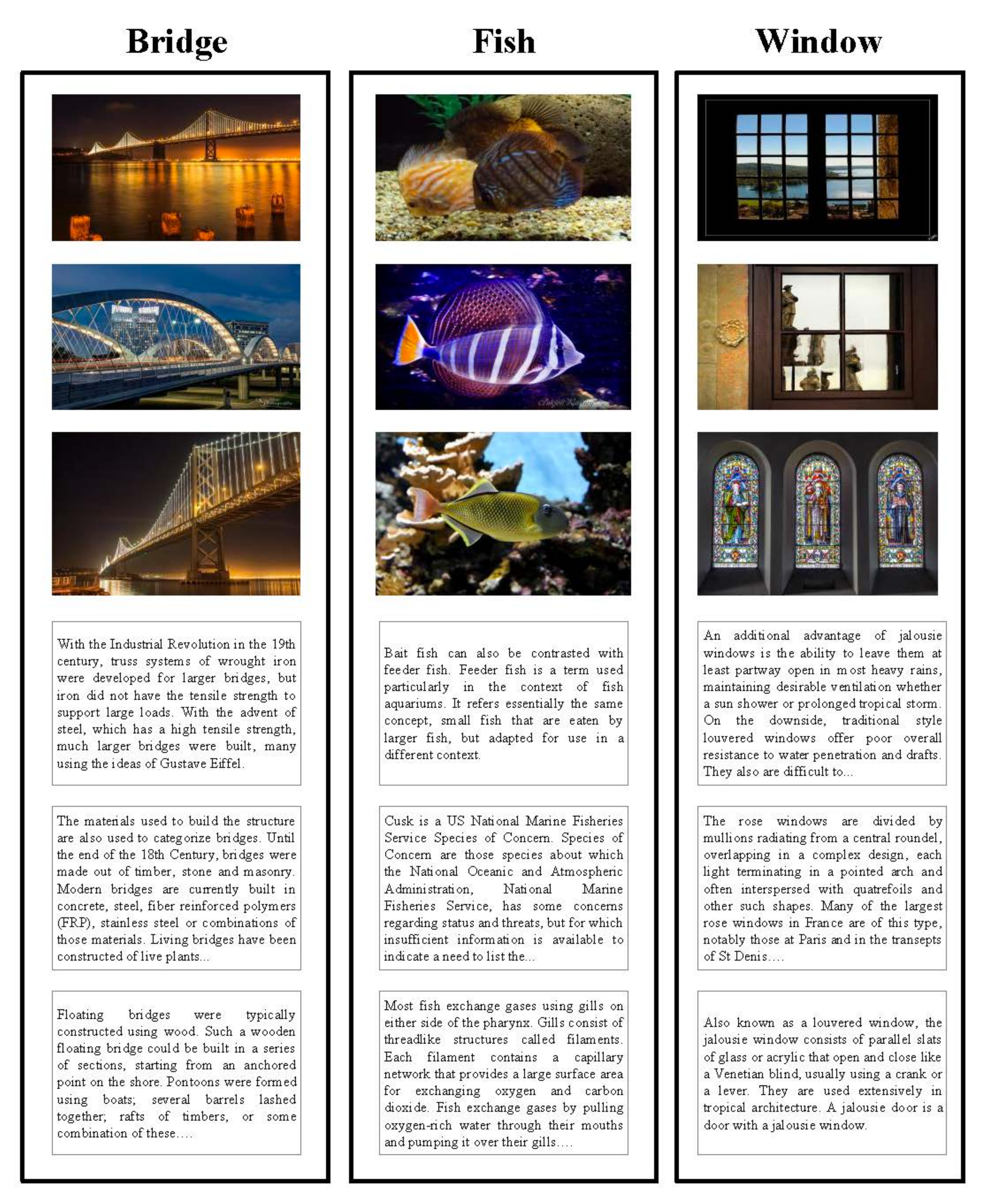}
	\caption{Image and text examples of 3 categories from XMediaNet dataset, which are bridge, fish and window.}
	\label{fig_wiki_ex}
\end{figure}

\begin{itemize}
	\item {  \textbf{XMediaNet dataset}} is a new large-scale dataset constructed by ourselves. It contains 5 media types, including text, image, audio, video and 3D model. Up to 200 independent categories are selected from the WordNet\footnote{http://wordnet.princeton.edu/}, including 47 animal species and 153 artifact species. In this paper, we use image and text data in XMediaNet dataset to conduct the experiments. There are both 40,000 instances for images and texts. The images are all crawled from Flickr\footnote{http://www.flickr.com}, while the texts are the paragraphs of the corresponding introductions in Wikipedia website. In the experiments, XMediaNet dataset is divided into three parts. Training set has 32,000 pairs, while validation set and test set both have 4,000 pairs. Some examples of this dataset are shown in Figure \ref{fig_wiki_ex}.
	
	\item { \textbf{Wikipedia dataset}} \cite{RasiwasiaMM10SemanticCCA} has 2,866 image/text pairs labelled by 10 categories, including history, biology and so on. For fair comparison, the dataset is divided into 3 subsets, which are training set with 2,173 pairs, testing set with 462 pairs and validation set with 231 pairs, following \cite{DBLP:conf/ijcai/PengHQ16,feng12014cross}.
	
	\item {  \textbf{Pascal Sentence dataset}} \cite{rashtchian2010collecting} is generated from 2008 PASCAL development kit, consisting of 1,000 images with 20 categories. Each image is described by 5 sentences, which are treated as a document. We divided this dataset into 3 subsets like Wikipedia dataset also following \cite{DBLP:conf/ijcai/PengHQ16,feng12014cross}, namely 800 pairs for training, 100 pairs for validation and 100 pairs for testing.

\end{itemize}

\begin{table*}[htb]
	\caption{The MAP scores of Cross-modal Retrieval for our CM-GANs approach and 10 compared methods on \textbf{XMediaNet} dataset.}
	\begin{center}
		\scalebox{1.0}{
			\begin{tabular}{|c|c|c|c|c|c|c|} 
				\hline
				\multirow{2}{*}{Method} & \multicolumn{3}{c|}{Bi-modal retrieval} & \multicolumn{3}{c|}{All-modal retrieval} \\
				\cline{2-7}
				& Image$\rightarrow$Text & Text$\rightarrow$Image & Average & Image$\rightarrow$All & Text$\rightarrow$All & Average\\
				\hline
				\textbf{Our CM-GANs Approach} & \textbf{0.567} & \textbf{0.551} & \textbf{0.559} & \textbf{0.581} & \textbf{0.590} & \textbf{0.586} \\
				CCL \cite{peng2017ccl} & 0.537 & 0.528 & 0.533  & 0.552 & 0.578 & 0.565\\
				CMDN \cite{DBLP:conf/ijcai/PengHQ16} & 0.485 & 0.516 & 0.501 & 0.504 & 0.563 & 0.534 \\
				Deep-SM \cite{DBLP:journals/tcyb/WeiZLWLZY17} & 0.399 & 0.342 & 0.371 & 0.351 & 0.338 & 0.345 \\
				LGCFL \cite{DBLP:journals/tmm/KangXLXP15} & 0.441 & 0.509 & 0.475 & 0.314 & 0.544 & 0.429\\
				JRL \cite{ZhaiTCSVT2014JRL} & 0.488 & 0.405& 0.447 & 0.508 & 0.505 & 0.507\\
				DCCA \cite{DBLP:conf/cvpr/YanM15} & 0.425 & 0.433& 0.429 & 0.433 & 0.475 & 0.454\\
				Corr-AE \cite{feng12014cross} & 0.469 & 0.507& 0.488 & 0.342 & 0.314 & 0.328\\
				KCCA \cite{DBLP:journals/neco/HardoonSS04} & 0.252 & 0.270& 0.261 & 0.299 & 0.186 & 0.243\\
				CFA \cite{LiMM03CFA} & 0.252 & 0.400& 0.326 & 0.318 & 0.207 & 0.263\\
				CCA \cite{HotelingBiometrika36RelationBetweenTwoVariates} & 0.212 & 0.217& 0.215  & 0.254 & 0.252 & 0.253\\
				\hline
				
			\end{tabular} 
		}
	\end{center}
	\label{table:xmn}
\end{table*}

\begin{table*}[htb]
	\caption{The MAP scores of Cross-modal Retrieval for our CM-GANs approach and 10 compared methods on \textbf{Pascal Sentence} dataset.}
	\begin{center}
		\scalebox{1.0}{
			\begin{tabular}{|c|c|c|c|c|c|c|} 
				\hline
				\multirow{2}{*}{Method}  & \multicolumn{3}{c|}{Bi-modal retrieval} & \multicolumn{3}{c|}{All-modal retrieval}\\
				\cline{2-7}
				& Image$\rightarrow$Text & Text$\rightarrow$Image & Average & Image$\rightarrow$All & Text$\rightarrow$All & Average\\
				\hline
				\textbf{Our CM-GANs Approach} & \textbf{0.603} & \textbf{0.604} & \textbf{0.604} & \textbf{0.584} & \textbf{0.698} & \textbf{0.641}\\
				CCL \cite{peng2017ccl} & 0.576 & 0.561 & 0.569  & 0.575 & 0.632 & 0.604\\
				CMDN \cite{DBLP:conf/ijcai/PengHQ16} & 0.544 & 0.526 & 0.535  & 0.496 & 0.627 & 0.562\\
				Deep-SM \cite{DBLP:journals/tcyb/WeiZLWLZY17} & 0.560 & 0.539 & 0.550  & 0.555 & 0.653 & 0.604\\
				LGCFL \cite{DBLP:journals/tmm/KangXLXP15} & 0.539 & 0.503 & 0.521 & 0.385 & 0.420 & 0.403\\
				JRL \cite{ZhaiTCSVT2014JRL} & 0.563 & 0.505& 0.534 & 0.561 & 0.631 & 0.596\\
				DCCA \cite{DBLP:conf/cvpr/YanM15} & 0.568 & 0.509& 0.539 & 0.556 & 0.653 & 0.605\\
				Corr-AE \cite{feng12014cross} & 0.532 & 0.521& 0.527 & 0.489 & 0.534 & 0.512\\
				KCCA \cite{DBLP:journals/neco/HardoonSS04} & 0.488 & 0.446& 0.467 & 0.346 & 0.429 & 0.388\\
				CFA \cite{LiMM03CFA} & 0.476 & 0.470& 0.473 & 0.470 & 0.497 & 0.484\\
				CCA \cite{HotelingBiometrika36RelationBetweenTwoVariates} & 0.203 & 0.208& 0.206  & 0.238 & 0.301 & 0.270\\
				\hline
				
			\end{tabular} 
		}
	\end{center}
	\label{table:pas}
\end{table*}

\begin{table*}[htb]
	\caption{The MAP scores of Cross-modal Retrieval for our CM-GANs approach and 10 compared methods on \textbf{Wikipedia} dataset.}
	\begin{center}
		\scalebox{1.0}{
			\begin{tabular}{|c|c|c|c|c|c|c|} 
				\hline
				\multirow{2}{*}{Method}  & \multicolumn{3}{c|}{Bi-modal retrieval} & \multicolumn{3}{c|}{All-modal retrieval}\\
				\cline{2-7}
				& Image$\rightarrow$Text & Text$\rightarrow$Image & Average & Image$\rightarrow$All & Text$\rightarrow$All & Average\\
				\hline
				\textbf{Our CM-GANs Approach} & \textbf{0.521} & \textbf{0.466} & \textbf{0.494} & \textbf{0.434} & \textbf{0.661} & \textbf{0.548}\\
				CCL \cite{peng2017ccl} & 0.505 & 0.457 & 0.481 & 0.422 & 0.652 & 0.537 \\
				CMDN \cite{DBLP:conf/ijcai/PengHQ16} & 0.487 & 0.427 & 0.457  & 0.407 & 0.611 & 0.509\\
				Deep-SM \cite{DBLP:journals/tcyb/WeiZLWLZY17} & 0.478 & 0.422 & 0.450  & 0.391 & 0.597 & 0.494\\
				LGCFL \cite{DBLP:journals/tmm/KangXLXP15} & 0.466 & 0.431 & 0.449 & 0.392 & 0.598 & 0.495\\
				JRL \cite{ZhaiTCSVT2014JRL} & 0.479 & 0.428& 0.454 & 0.404 & 0.595 & 0.500\\
				DCCA \cite{DBLP:conf/cvpr/YanM15} & 0.445 & 0.399& 0.422 & 0.371 & 0.560 & 0.466\\
				Corr-AE \cite{feng12014cross} & 0.442 & 0.429& 0.436 & 0.397 & 0.608 & 0.494\\
				KCCA \cite{DBLP:journals/neco/HardoonSS04} & 0.438 & 0.389& 0.414 & 0.354 & 0.518 & 0.436\\
				CFA \cite{LiMM03CFA} & 0.319 & 0.316& 0.318 & 0.279 & 0.341 & 0.310\\
				CCA \cite{HotelingBiometrika36RelationBetweenTwoVariates} & 0.298 & 0.273& 0.286 & 0.268 & 0.370 & 0.319 \\
				\hline
				
			\end{tabular} 
		}
	\end{center}
	\label{table:wiki}
\end{table*}

\subsection{Evaluation Metric}

The heterogeneous data can be correlated with the learned common representation by similarity metric. To comprehensively evaluate the performance of cross-modal correlation, we preform cross-modal retrieval with two kinds of retrieval tasks on 3 datasets, namely \textit{bi-modal retrieval} and \textit{all-modal retrieval}, which are defined as follows.

\subsubsection{\textbf{Bi-modal retrieval}} To perform retrieval between different modalities with the following two sub-tasks.

\begin{itemize}
	\item {Image retrieve text} (image$\to$text): Taking images as queries, to retrieve text instances in the testing set by calculated cross-modality similarity.
	\item {Text retrieve image} (text$\to$image): Taking texts as queries, to retrieve image instances in the testing set by calculated cross-modality similarity.
\end{itemize}

\subsubsection{\textbf{All-modal retrieval}} To perform retrieval among all modalities with the following two sub-tasks.

\begin{itemize}
	\item {Image retrieve all modalities} (image$\to$all): Taking images as queries, to retrieve both text and image instances in the testing set by calculated cross-modality similarity.
	\item {Text retrieve all modalities} (text$\to$all): Taking texts as queries, to retrieve both text and image instances in the testing set by calculated cross-modality similarity.
\end{itemize}

It should be noted that all the compared methods adopt the same CNN features for both image and text extracted from the CNN architectures used in our approach for fair comparison. Specifically, we extract CNN feature for image from the fc7 layer in 19-layer VGGNet \cite{DBLP:journals/corr/SimonyanZ14a}, and CNN feature for text from Word CNN with the same configuration of \cite{DBLP:conf/emnlp/Kim14}. Besides, we use the source codes released by their authors to evaluate the compared methods fairly with the following steps: (1) Common representation learning with the training data to learn the projections or deep models. (2) Converting the testing data into the common representation by the learned projections or deep models. (3) Computing cross-modal similarity with cosine distance to perform cross-modal retrieval.

For the evaluation metric, we calculate mean average precision (MAP) score for all returned results on all the 3 datasets. First, the Average Precision (AP) is calculated for each query as follows:
\begin{align}
AP= \frac{1}{R}\sum_{k=1}^{n}\frac{R_{k}}{k}\times rel_{k},
\end{align}
where $n$ denotes the total number of instance in testing set consisting of $R$ relevant instances. The top $k$ returned results contain $R_k$ relevant instances. If the $k$-th returned result is relevant, $rel_k$ is set to be 1, otherwise, $rel_k$ is set to be 0. Then, the mean value of calculated AP on each query is formed as MAP, which joint considers the ranking information and precision and is widely used in cross-modal retrieval task.

\subsection{Compared Methods}

To verify the effectiveness of our proposed CM-GANs approach, we compare 10 state-of-the-art methods in the experiments, including 5 traditional cross-modal retrieval methods, namely CCA \cite{HotelingBiometrika36RelationBetweenTwoVariates}, CFA \cite{LiMM03CFA}, KCCA \cite{DBLP:journals/neco/HardoonSS04}, JRL \cite{ZhaiTCSVT2014JRL} and LGCFL \cite{DBLP:journals/tmm/KangXLXP15}, as well as 5 deep learning based methods, namely Corr-AE \cite{feng12014cross}, DCCA \cite{DBLP:conf/cvpr/YanM15}, CMDN \cite{DBLP:conf/ijcai/PengHQ16}, Deep-SM \cite{DBLP:journals/tcyb/WeiZLWLZY17} and CCL \cite{peng2017ccl}. We briefly introduce these compared methods in the following paragraphs.

\begin{itemize}
	\item {\bf CCA} \cite{HotelingBiometrika36RelationBetweenTwoVariates} learns projection matrices to map the features of different modalities into one common space by maximizing the correlation on them.
	
	\item {\bf CFA} \cite{LiMM03CFA} minimizes the Frobenius norm and projects  the data of different modalities into one common space.
	
	\item {\bf KCCA} \cite{DBLP:journals/neco/HardoonSS04} adopts kernel function to extend CCA for the common space learning. In the experiments, Gaussian kernel is used as the kernel function.
	
	\item {\bf JRL} \cite{ZhaiTCSVT2014JRL} adopts semi-supervised regularization as well as sparse regularization to learn the common space with semantic information.
	
	\item {\bf LGCFL} \cite{DBLP:journals/tmm/KangXLXP15} uses a local group based priori to exploit popular block based features and jointly learns basis matrices for different modalities.
	
	\item {\bf Corr-AE} \cite{feng12014cross} jointly models the correlation and reconstruction learning error with two subnetworks linked at the code layer, which has two extensions, and the best results of these models for fair comparison is reported in the experiments.
	
	\item {\bf DCCA} \cite{DBLP:conf/cvpr/YanM15} adopts the similar objective function with CCA on the top of two separate subnetworks to maximize the correlation between them.
	
	\item {\bf CMDN} \cite{DBLP:conf/ijcai/PengHQ16} jointly models the intra-modality and inter-modality correlation in both separate representation and common representation learning stages with multiple deep networks.
	
	\item {\bf Deep-SM} \cite{DBLP:journals/tcyb/WeiZLWLZY17} performs deep semantic matching to exploit the strong representation learning ability of convolutional neural network for image.
	
	\item {\bf CCL} \cite{peng2017ccl} fully explores both intra-modality and inter-modality correlation simultaneously with multi-grained and multi-task learning.
	
\end{itemize}

\begin{figure*}[!t]
	\centering
	\includegraphics[width=0.93\textwidth]{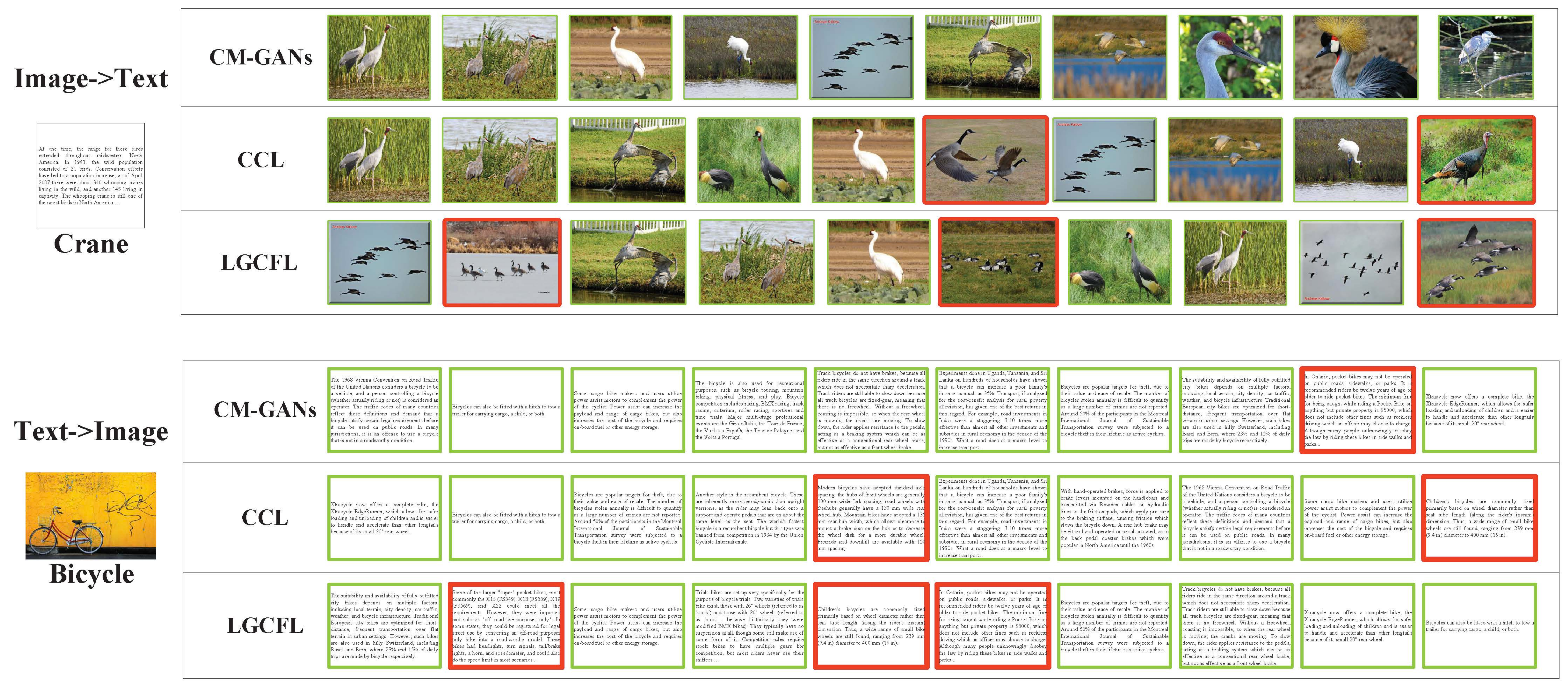}
	\caption{Examples of the bi-modal retrieval results on XMediaNet dataset by our proposed CM-GANs approach as well as the best compared deep learning based method CCL \cite{peng2017ccl} and the best compared traditional method LGCFL \cite{DBLP:journals/tmm/KangXLXP15}. The retrieval results with green borders are correct, while those with red borders are wrong in these examples.
	}
	\label{fig_xmn_res}
\end{figure*}

\subsection{Comparisons with 10 State-of-the-art Methods}

In this subsection, we compare the cross-modal retrieval accuracy to evaluate the effectiveness of the learned common representation on both our proposed approach as well as the state-of-the-art compared methods. The experimental results are shown in Tables \ref{table:xmn}, \ref{table:pas} and \ref{table:wiki}, including the MAP scores of both bi-modal retrieval and all-modal retrieval on 3 datasets, from which we can observe that our proposed CM-GANs approach achieves the best retrieval accuracy among all the compared methods. On our constructed large-scale XMediaNet dataset as shown in Table \ref{table:xmn}, the average MAP score of bi-modal retrieval has been improved from 0.533 to 0.559, while our proposed approach also makes improvement on all-modal retrieval.
Among the compared methods, most deep learning based methods have better performance than the traditional methods, where CCL achieves the best accuracy in the compared methods, and some traditional methods also get benefits from the CNN feature leading to a close accuracy with the deep learning based methods, such as LGCFL and JRL, which are the two best compared traditional methods.

Besides, on Pascal Sentence and Wikipedia datasets, we can also observe similar trends on the results of bi-modal retrieval and all-modal retrieval, which are shown in Tables \ref{table:pas} and \ref{table:wiki}. Our proposed approach outperforms all the compared methods and achieves great improvement on the MAP scores. 
For intuitive comparison, we have shown some bi-modal retrieval results in Figure \ref{fig_xmn_res} on our constructed large-scale XMediaNet dataset.

 \begin{figure*}[!t]
	\centering
	\includegraphics[width=1.0\textwidth]{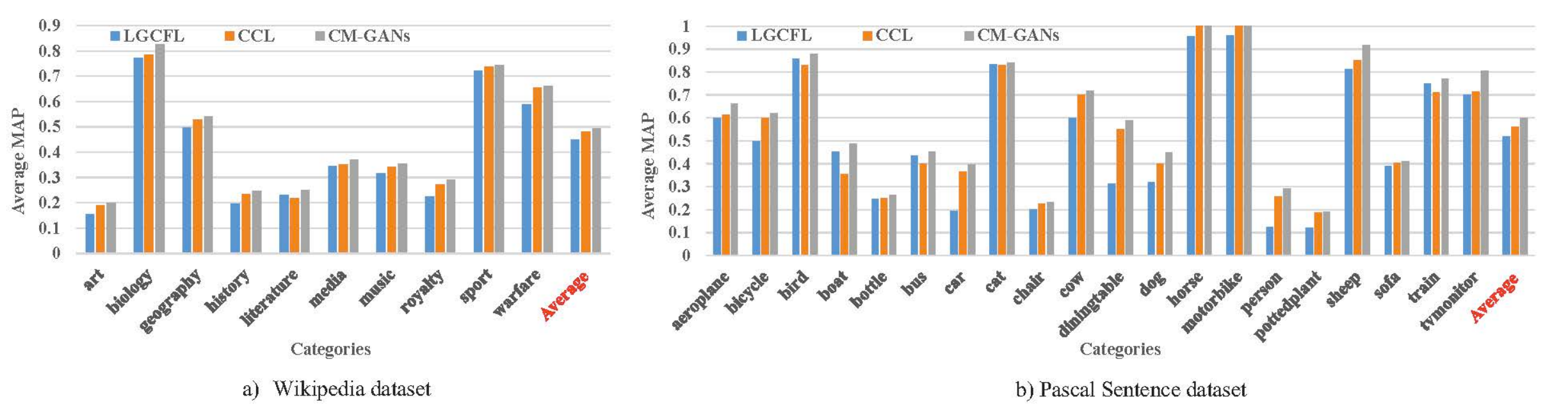}
	\caption{The respective result of each category on our proposed approach as well as the compared CCL \cite{peng2017ccl} and LGCFL \cite{DBLP:journals/tmm/KangXLXP15} methods, in Wikipedia dataset and Pascal Sentence dataset. We can see that the retrieval accuracies differ greatly between different modalities. Some categories with high-level semantics, such as ``art'' and ``history'' in Wikipedia dataset, or with relatively small object such as ``bottle'' and ``potted plant'' in Pascal Sentence dataset, may lead to confusions when performing cross-modal retrieval.
	}
	\setlength{\abovecaptionskip}{-1.5cm}
	\label{fig_cat}
\end{figure*}

\subsection{Experimental Analysis}

The in-depth experimental analysis is presented in this subsection of our proposed approach and the compared state-of-the-art methods. We also give some failure analysis on our proposed approach for further discussion.

First, for compared deep learning based methods, DCCA, Corr-AE and Deep-SM all have similar network structures that consist of two subnetworks. Corr-AE jointly models the cross-modal correlation learning error as well as the reconstruction error. Although DCCA only maximizes the correlation on the top of two subnetworks, it utilizes the strong representation learning ability of convolutional neural network to reach roughly the same accuracies with Corr-AE. While Deep-SM further integrates semantic category information to achieve better accuracy. Besides, both CMDN and CCL contain multiple deep networks to consider intra-modality and inter-modality correlation in a multi-level framework, which makes them outperform the other methods. While CCL further exploits the fine-grained information as well as adopts multi-task learning strategy to get the best accuracy among the compared methods. Then, for the traditional methods, although their performance benefits from the deep feature, most of them are still limited in the traditional framework and get poor accuracies such as CCA and CFA. KCCA, as an extension of CCA, achieves better accuracy because of the kernel function to model the nonlinear correlation. Besides, JRL and LGCFL have the best retrieval accuracies among the traditional methods, and even outperform some deep learning based methods, for the fact that JRL adopts semi-supervised and sparse regularization, while LGCFL uses a local group based priori to take the advantage of popular block based features.

Compared with the above state-of-the-art methods, our proposed CM-GANs approach clearly keeps the advantages as shown in Tables \ref{table:xmn}, \ref{table:pas} and \ref{table:wiki} for the 3 reasons as follows: 
(1) Cross-modal GANs architecture fully models the joint distribution over the data of different modalities with cross-modal adversarial training process.
(2) Cross-modal convolutional autoencoders with weight-sharing and semantic constraints as the generative model fit the joint distribution by exploiting both inter-modality and intra-modality correlation.
(3) Inter-modality and intra-modality discrimination in the discriminative model strengthens the generative model.

Then, for the failure analysis, Figures \ref{fig_xmn_res} and \ref{fig_cat} show the retrieval results in XMediaNet dataset and the MAP score of each category in Wikipedia and Pascal Sentence datasets. From Figure \ref{fig_xmn_res}, we can observe that the failure cases are mostly caused by the small variance between image instances or the confusion in text instances among different categories, which leads to wrong retrieval results. But it should be noted that the number of failure cases can be effectively reduced with our proposed approach comparing with CCL as the best compared deep learning based method as well as LGCFL as the best compared traditional method. Besides, as shown in Figure \ref{fig_cat}, the retrieval accuracies of different categories differ from each other greatly. Some categories with high-level semantics, such as ``art'' and ``history'' in Wikipedia dataset, or with relatively small objects such as ``bottle'' and ``potted plant'' in Pascal Sentence dataset, may lead to confusions when performing cross-modal retrieval. However, our proposed approach still achieves the best retrieval accuracies on most categories compared with CCL and LGCFL, which indicates the effectiveness of our approach.

\begin{table*}[htb]
	\caption{Baseline experiments on \textbf{Performance of generative model}, where \textbf{CM-GANs without ws\&sc} means none constraint is adopted in generative model, while \textbf{CM-GANs with ws} and \textbf{CM-GANs with sc} mean either weigh-sharing or semantic constraint is adopted.}
	\begin{center}
		\scalebox{1.0}{
			\begin{tabular}{|c|c|c|c|c|} 
				\hline
				\multirow{2}{*}{Dataset}&
				\multirow{2}{*}{Method} & \multicolumn{3}{c|}{MAP scores} \\
				\cline{3-5}
				& & Image$\rightarrow$Text & Text$\rightarrow$Image & Average \\
				\hline

				\multirow{4}{*}{\begin{tabular}{c} XMediaNet dataset \end{tabular}} 
				&  \textbf{Our CM-GANs Approach} & \textbf{0.567} & \textbf{0.551} & \textbf{0.559}  \\
				
				&   \multirow{1}{*}{\begin{tabular}{c}CM-GANs without ws\&sc\end{tabular}} & \multirow{1}{*}{0.530} & \multirow{1}{*}{0.524}& \multirow{1}{*}{0.527} \\
				& \multirow{1}{*}{\begin{tabular}{c}CM-GANs with ws\end{tabular}} & \multirow{1}{*}{0.548} & \multirow{1}{*}{0.544}& \multirow{1}{*}{0.546} \\
				& \multirow{1}{*}{\begin{tabular}{c}CM-GANs with sc\end{tabular}} & \multirow{1}{*}{0.536} & \multirow{1}{*}{0.529}& \multirow{1}{*}{0.533} \\
				\hline

				\multirow{4}{*}{\begin{tabular}{c} Pascal Sentence \\ dataset \end{tabular}} 
				&  \textbf{Our CM-GANs Approach} & \textbf{0.603} & \textbf{0.604} & \textbf{0.604} \\
				
				&   \multirow{1}{*}{\begin{tabular}{c}CM-GANs without ws\&sc\end{tabular}} & \multirow{1}{*}{0.562} & \multirow{1}{*}{0.557}& \multirow{1}{*}{0.560} \\
				& \multirow{1}{*}{\begin{tabular}{c}CM-GANs with ws\end{tabular}} & \multirow{1}{*}{0.585} & \multirow{1}{*}{0.586}& \multirow{1}{*}{0.585} \\
				& \multirow{1}{*}{\begin{tabular}{c}CM-GANs with sc\end{tabular}} & \multirow{1}{*}{0.566} & \multirow{1}{*}{0.570}& \multirow{1}{*}{0.568} \\
				\hline
				
				\multirow{4}{*}{\begin{tabular}{c} Wikipedia dataset \end{tabular}}
				& \textbf{Our CM-GANs Approach} & \textbf{0.521} & \textbf{0.466} & \textbf{0.494} \\
				
				&   \multirow{1}{*}{\begin{tabular}{c}CM-GANs without ws\&sc\end{tabular}} & \multirow{1}{*}{0.489} & \multirow{1}{*}{0.436}& \multirow{1}{*}{0.463}\\
				& \multirow{1}{*}{\begin{tabular}{c}CM-GANs with ws\end{tabular}} & \multirow{1}{*}{0.502} & \multirow{1}{*}{0.439}& \multirow{1}{*}{0.470} \\
				& \multirow{1}{*}{\begin{tabular}{c}CM-GANs with sc\end{tabular}} & \multirow{1}{*}{0.494} & \multirow{1}{*}{0.438}& \multirow{1}{*}{0.466} \\
				\hline
				
			\end{tabular} 
		}
	\end{center}
	\label{table:Baseline1}
\end{table*}

\begin{table*}[htb]
	\caption{Baseline experiments on Performance of \textbf{discriminative model}, where \textbf{CM-GANs only inter} means that only the inter-modality discrimination is adopted.}
	\begin{center}
		\scalebox{1.0}{
			\begin{tabular}{|c|c|c|c|c|} 
				\hline
				\multirow{2}{*}{Dataset}&
				\multirow{2}{*}{Method} & \multicolumn{3}{c|}{MAP scores} \\
				\cline{3-5}
				& & Image$\rightarrow$Text & Text$\rightarrow$Image & Average \\
				\hline
				
				\multirow{2}{*}{\begin{tabular}{c} XMediaNet dataset \end{tabular}} 
				&  \textbf{Our CM-GANs Approach} & \textbf{0.567} & \textbf{0.551} & \textbf{0.559}  \\
				&   \multirow{1}{*}{\begin{tabular}{c}CM-GANs only inter\end{tabular}} & \multirow{1}{*}{0.529} & \multirow{1}{*}{0.524}& \multirow{1}{*}{0.527} \\
				\hline
				
				\multirow{2}{*}{\begin{tabular}{c} Pascal Sentence \\ dataset \end{tabular}} 
				&  \textbf{Our CM-GANs Approach} & \textbf{0.603} & \textbf{0.604} & \textbf{0.604} \\
				
				&   \multirow{1}{*}{\begin{tabular}{c}CM-GANs only inter\end{tabular}} & \multirow{1}{*}{0.576} & \multirow{1}{*}{0.577}& \multirow{1}{*}{0.577} \\
				\hline

				\multirow{2}{*}{\begin{tabular}{c} Wikipedia dataset \end{tabular}}
				& \textbf{Our CM-GANs Approach} & \textbf{0.521} & \textbf{0.466} & \textbf{0.494} \\
				
				&   \multirow{1}{*}{\begin{tabular}{c}CM-GANs only inter\end{tabular}} & \multirow{1}{*}{0.506} & \multirow{1}{*}{0.442}& \multirow{1}{*}{0.474} \\
				\hline
				
			\end{tabular} 
		}
	\end{center}
	\label{table:Baseline2}
\end{table*}

\begin{table*}[htb]
	\caption{Baseline experiments on Performance of \textbf{adversarial training}, where \textbf{CM-GANs-CAE} means to train the generative model without adversarial training process.}
	\begin{center}
		\scalebox{1.0}{
			\begin{tabular}{|c|c|c|c|c|} 
				\hline
				\multirow{2}{*}{Dataset}&
				\multirow{2}{*}{Method} & \multicolumn{3}{c|}{MAP scores} \\
				\cline{3-5}
				& & Image$\rightarrow$Text & Text$\rightarrow$Image & Average \\
				\hline
				
				\multirow{2}{*}{\begin{tabular}{c} XMediaNet dataset \end{tabular}} 
				&  \textbf{Our CM-GANs Approach} & \textbf{0.567} & \textbf{0.551} & \textbf{0.559}  \\
				&   \multirow{1}{*}{\begin{tabular}{c}CM-GANs-CAE\end{tabular}} & \multirow{1}{*}{0.491} & \multirow{1}{*}{0.511}& \multirow{1}{*}{0.501} \\
				\hline
				
				\multirow{2}{*}{\begin{tabular}{c} Pascal Sentence \\ dataset \end{tabular}} 
				&  \textbf{Our CM-GANs Approach} & \textbf{0.603} & \textbf{0.604} & \textbf{0.604} \\
				
				&   \multirow{1}{*}{\begin{tabular}{c}CM-GANs-CAE\end{tabular}} & \multirow{1}{*}{0.563} & \multirow{1}{*}{0.545}& \multirow{1}{*}{0.554} \\
				\hline

				\multirow{2}{*}{\begin{tabular}{c} Wikipedia dataset \end{tabular}}
				& \textbf{Our CM-GANs Approach} & \textbf{0.521} & \textbf{0.466} & \textbf{0.494} \\
				
				&   \multirow{1}{*}{\begin{tabular}{c}CM-GANs-CAE\end{tabular}} & \multirow{1}{*}{0.460} & \multirow{1}{*}{0.436}& \multirow{1}{*}{0.448} \\
				\hline
				
			\end{tabular} 
		}
	\end{center}
	\label{table:Baseline3}
\end{table*}

\subsection{Baseline Comparisons}

To verify the effectiveness of each part in our proposed CM-GANs approach, three kinds of baseline experiments are conducted, and Tables \ref{table:Baseline1}, \ref{table:Baseline2} and \ref{table:Baseline3} show the comparison of our proposed approach with the baseline approaches. The detailed analysis is given in the following paragraphs.

\subsubsection{Performance of generative model}

We have constructed the cross-modal convolutional autoencoders with both weight-sharing and semantic constraints in the generative model, as mentioned in Section III.B. To demonstrate the separate contribution on each of them, we conduct 3 sets of baseline experiments, where ``ws'' denotes the weight-sharing constraint and ``sc'' denotes the semantic constraints. Thus, ``CM-GANs without ws\&sc'' means that none of these two constraints is adopted, and ``CM-GANs with ws'' and ``CM-GANs with sc'' means one of them is adopted. 

As shown in Table \ref{table:Baseline1}, these two components in the generative model have similar contributions on the accuracies for final cross-modal retrieval results, while weight-sharing constraint can effectively handle the cross-modal correlation and semantic constraints can preserve the semantic consistency between different modalities. Finally, both of them can mutually boost the common representation learning.

\subsubsection{Performance of discriminative model}

There are two kinds of discriminative models to simultaneously conduct the inter-modality discrimination and intra-modality discrimination. It should be noted that the inter-modality discrimination is indispensable for the cross-modal correlation learning. Therefore, we only conduct the baseline experiment on the effectiveness of intra-modality discrimination ``CM-GANs only inter''. 

As shown in Table \ref{table:Baseline2}, CM-GANs achieves the improvement on the average MAP score of bi-modal retrieval in 3 datasets. This indicates that the intra-modality discrimination plays a complementary role with inter-modality discrimination, which can preserve the semantic consistency within each modality by discriminating the generated reconstruction representation with the original representation.

\subsubsection{Performance of adversarial training}

We aim to verify the effectiveness of the adversarial training process. In our proposed approach, the generative model can be trained solely without discriminative model, by adopting the reconstruction learning error on the top of two decoders for each modality as well as weight-sharing and semantic constraints. This baseline approach is denoted as ``CM-GANs-CAE''.

From the results in Table \ref{table:Baseline3}, we can observe that CM-GANs obtains higher accuracy than CM-GANs-CAE on the average MAP score of bi-modal retrieval in 3 datasets. It demonstrates that the adversarial training process can effectively boost the cross-modal correlation learning to improve the performance of cross-modal retrieval.

The above baseline results have verified the separate contribution of each component in our proposed CM-GANs approach with the following 3 aspects: 
(1) Weight-sharing and semantic constraints can exploit the cross-modal correlation and semantic information between different modalities.
(2) Intra-modality discrimination can model semantic information within each modality to make complementary contribution to inter-modality discrimination.
(3) Cross-modal adversarial training can fully capture the cross-modal joint distribution to learn more discriminative common representation.

\section{Conclusion}

In this paper, we have proposed Cross-modal Generative Adversarial Networks (CM-GANs) to handle the heterogeneous gap to learn common representation for different modalities. First, cross-modal GANs architecture is proposed to fit the joint distribution over the data of different modalities with a minimax game. Second, cross-modal convolutional autoencoders are proposed with both weight-sharing and semantic constraints to model the cross-modal semantic correlation between different modalities. Third, a cross-modal adversarial mechanism is designed with two kinds of discriminative models to simultaneously conduct inter-modality and intra-modality discrimination for mutually boosting to learn more discriminative common representation. We conduct cross-modal retrieval to verify the effectiveness of the learned common representation, and our proposed approach outperforms 10 state-of-the-art methods on widely-used Wikipedia and Pascal Sentence datasets as well as our constructed large-scale XMediaNet dataset in the experiments.

For the future work, we attempt to further model the joint distribution over the data of more modalities, such as video, audio. Besides, we attempt to make the best of large-scale unlabeled data to perform unsupervised training for marching toward the practical application.


%

%


\ifCLASSOPTIONcaptionsoff
  \newpage
\fi



\bibliographystyle{IEEEtran}
\bibliography{ijcai16}
\end{document}